\documentclass[ceqn,10pt]{wlscirep}
\usepackage[utf8]{inputenc}
\usepackage[T1]{fontenc}
\usepackage{graphicx}
\usepackage{amsmath}
\usepackage{amssymb}
\usepackage{mathptmx}
\usepackage{orcidlink}

\title{New disordered anyon phase of doped \textcolor{black}{graphene} zigzag nanoribbon}

\author[1]{Young Heon Kim\orcidlink{0000-0003-0000-539X}}
\author[1]{Hye Jeong Lee\orcidlink{0000-0003-4535-9511}}
\author[2,3,4]{Hyun-Yong Lee\orcidlink{0000-0003-4774-5213}}
\author[1,*]{S.-R. Eric Yang\orcidlink{0000-0003-3377-1859}}
\affil[1]{Department of Physics, Korea University, Seoul 02855, Korea}
\affil[2]{Department of Applied Physics, Graduate School, Korea University, Sejong 30019, Korea}
\affil[3]{Division of Display and Semiconductor Physics, Korea University, Sejong 30019, Korea}
\affil[4]{Interdisciplinary Program in E$\cdot$ICT-Culture-Sports Convergence, Korea University, Sejong 30019, Korea}
\affil[*]{corresponding author: eyang812@gmail.com}

\begin{abstract}
We investigate interacting disordered zigzag nanoribbons at low doping\textcolor{black}{, using the Hubbard model to treat electron interactions within the density matrix renormalization group and Hartree-Fock method.}
Extra electrons that are inserted into an interacting disordered zigzag nanoribbon divide into anyons.
Furthermore, the
fractional charges form a new disordered anyon phase with a highly distorted edge spin density wave, containing numerous localized magnetic moments residing on the zigzag edges, thereby displaying spin-charge separation and a strong non-local correlation between the opposite zigzag edges. We make the following new predictions, which can be experimentally tested:
(1) In the low doping case \textcolor{black}{and weak disorder regime}, the soft gap in the tunneling density of states is replaced by a sharp peak at the midgap energy with two accompanying peaks. The $e^-/2$ fractional charges that reside on the boundary of the zigzag edges are responsible for these peaks. (2) We find that the midgap peak disappears as the doping concentration increases. The presence of $e^-/2$ fractional charges will be strongly supported by the detection of these peaks. Doped zigzag ribbons may also exhibit unusual transport, magnetic, and inter-edge tunneling properties.
\end{abstract}
\begin{document}

\flushbottom
\maketitle
%
%

\section*{Introduction}
Topological phases of matter are classified into two categories\cite{Wen11,Wen21}: symmetry-protected topological\,(SPT) phase and topologically ordered\,(TO) phase. The SPT phase is distinguished by short-ranged entanglement, whereas the TO phase is distinguished by long-range entanglement. Furthermore, the SPT phase protects boundary gapless states, and it cannot be adiabatically connected to a trivial product state under perturbations preserving a certain symmetry. In contrast, in the TO phase, the global pattern of entanglement causes topological ground state degeneracy, which is robust to local perturbation regardless of symmetry. Furthermore, the topological degeneracy generates a universal subleading term in the entanglement entropy, which is known as the topological entanglement entropy (TEE)\cite{Kitaev11, Levin11}. This entanglement entropy has been mainly used to detect the topological order\cite{Bal}. Topological order frequently results in topological excitations with fractional quantum numbers. Entanglement entropy may also show signs of topologically ordered insulators\cite{Haldane191}.

\begin{figure}[htpb]
\begin{center}
\includegraphics[width=0.6\textwidth]{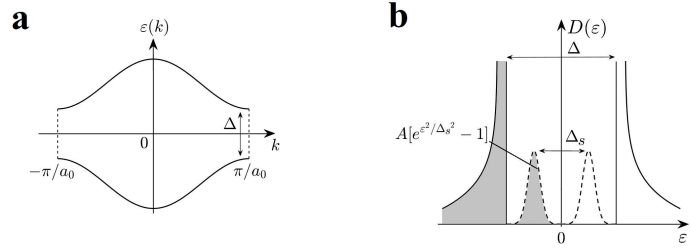}
\caption{\normalfont (\textbf{a}) Schematic band structure of a disorder-free interacting zigzag graphene nanoribbon at half-filling with a hard Mott gap $\Delta\sim 0.1t$, where the hopping parameter $t\sim 3$\;eV. Soliton zigzag edge states are near $k=\pm \pi/a_0$, and their charge is $e^-$ (the ribbon period is $a_0$). (\textbf{b}) Schematic density of states (DOS) of a half-filled disordered zigzag ribbon (dashed line). It decays exponentially with an energy scale of $\Delta_s$ (a soft gap), which decreases with increasing disorder strength\cite{Jeong11}. The van Hove singularities of the DOS at $\epsilon=\pm \Delta/2$, originating from   the band structure displayed in a, are reduced due to the formation of the gap states. These gap states are spin--split, and many of them are soliton states with fractional charge $e^-/2$. Note that the DOS has particle-hole symmetry after disorder averaging. The soft gap $\Delta_s$ protects the fractional charges from quantum fluctuations.}
\label{degGap}
\end{center}
\end{figure}

Recently, it was revealed that undoped interacting disordered  graphene\cite{Nov,Zhang,Neto} zigzag ribbons\cite{Fujita} are a new TO Mott--Anderson insulator displaying $e^-/2$ fractional charges,
spin-charge separation, and two degenerate ground states\cite{Jeong11,eeyang11}. The disorder is a singular perturbation that couples electrons on opposing zigzag edges, resulting in instantons. This effect converts zigzag ribbons from a STP to a TO phase and generates $e^-/2$  fractional charges on the opposite zigzag edges. These fractional charges are protected\cite{Girvin117} by an exponentially decaying soft gap\cite{Efros,Mac1} $\Delta_s$, as shown in Fig. \ref{degGap}. Furthermore, numerical work\cite{Kim11} showed that an interacting disordered zigzag nanoribbon has a finite TEE.

Recent advancements in fabrication methods have enabled the production of atomically precise graphene nanoribbons\cite{Cai2,Kolmer}.  But it is unclear how to unequivocally measure the presence of fractional charges. We believe that doped zigzag nanoribbons are ideal for observing exotic anyons with fractional charges. The properties of doped disordered zigzag ribbons, on the other hand, are largely unknown.
A doped ribbon is not expected to be a topologically ordered insulator because there is no hard gap (the density of states (DOS) at the Fermi is non-zero but small). However,
  the system is still an insulator with localized edge states near the Fermi energy, displaying doubly degenerate ground states. In the dilute limit the added fractional charges will still be well defined. Let us explain this, following  Ref.\cite{Girvin117}. These fractional charges are analogous to quasiparticles of the fractional quantum Hall effect's $1/m$ Laughlin state ($m$ is an odd integer). In such a system's low doped regime, the added electrons divide into fractional charges.  \textcolor{black}{Recent experimental works provide evidence for these anyons\cite{Nakamura01,Barto1}.}   Suppose one adds $\delta N$ electrons to such a state. In the dilute limit, each of these electrons fractionalizes into $m$ quasiparticles that are well separated from each other (the charge of a quasiparticle is $e^-/m$). The total energy of the new system is thus $E_-=E_m+\delta N m \Delta_-$, where $E_m$ is the ground state energy and $\Delta_{-}$ is the quasiparticle excitation energy. \textcolor{black}{Despite that the quasiparticles form quasi-degenerate states, the} excitation gap $\Delta$ and localization of quasiparticles protect fractional charges against quantum fluctuations~\textcolor{black}{\cite{eeyang11,Girvin117}}.

The role of fractional charges in low doped disordered systems is one of the fundamental questions in doped disordered zigzag ribbons. What exactly is the ground state? This concerns the applicability of mean field approaches to such a system: quantum fluctuations\cite{Girvin117} not included in the Hartree--Fock (HF) approximation may be significant because gap states are no longer empty.
Furthermore, in contrast to the uniform spin density of undoped ribbons, the ground state of a doped disorder-free ribbon exhibits an edge spin density wave. It is unknown how localization and charge quantization affect the nature of the ground state.
We use the density matrix renormalization group \,(DMRG) approach in the matrix product states (MPS) representation to investigate the ground state of a doped ribbon and the importance of quantum fluctuations beyond the HF approach. The MPS representation is a powerful tool for solving eigenvalue problems of quantum many-body systems\cite{White1992,Schollwock2011}.

Ribbons are in a new disordered anyon phase, according to our investigation of the low doping regime\cite{Lei13,Wilczek03}.
We discover that a low doped disordered zigzag ribbon contains a large number of anyons with a fractional charge (but as doping concentration increases they disappear).
They cause numerous magnetic domain walls and localized magnetic moments residing on the zigzag edges.
Also, objects that display spin-charge separation proliferate in this phase.
As a result, the ground state is drastically reorganized, with highly distorted edge charge and spin modulations, as well as non-local correlations between the left and right zigzag edges.
We will define this new phase as a disordered anyon phase because its electron and spin densities are highly inhomogeneous.  Furthermore, we make the following new experimentally testable predictions.
(1) The disordered anyon phase has an unusual shape of tunneling density of states (TDOS), depending on the number of extra electrons \textcolor{black}{(for experimental measurement of a soft gap in the TDOS, see, for example, Refs. \cite{Ashoori01,JPEisen})}.
The TDOS has one sharp peak at the midgap energy and two other peaks, one on each side of the sharp peak at the midgap energy, at the low doping limit. (2) However, the midgap peak disappears as the doping concentration increases. The detection of these peaks will provide strong evidence for the presence of $e^-/2$ fractional charges. Furthermore, our findings indicate that doped zigzag ribbons may have unusual transport, magnetic, and inter-edge tunneling properties. \textcolor{black}{Theoretical calculations\cite{Pisa1} of disorder-free zigzag ribbons show \textcolor{black}{that} antiferromagnetism is favored over ferromagnetism  for ribbon widths $< 100$\AA. In the presence of disorder, the new disordered anyon phase is expected for these width values.} 
\section*{Results}
\subsection*{Model}
\begin{figure}[htpb]
\begin{center}
\includegraphics[width=0.9\textwidth]{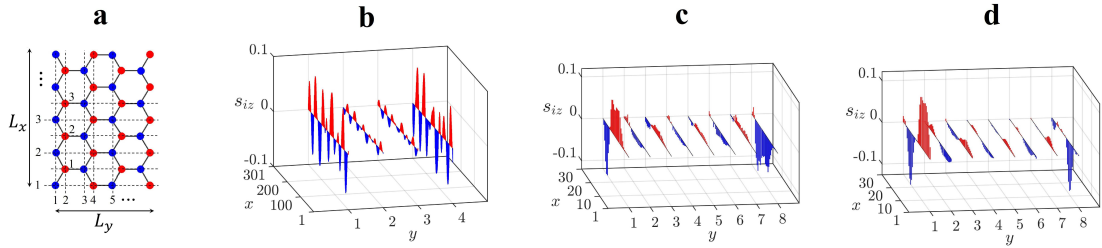}
\caption{\normalfont (\textbf{a}) Vertical and horizontal lines of carbon atoms are numbered. All lengths and widths in this paper are measured in the number of these lines. (\textbf{b})  Site spin values $ s_{iz}$ of a disorder-free  doped zigzag ribbon. This state is generated from  $N_e=N_s+20$ ($\delta N/N_s=0.017$), $L_x=301$, $L_y=4$, and $U=t$. (\textbf{c}) DMRG result of the ground state site spins $s_{iz}$ at half-filling for $U=t$, $n_{imp}=1$, and $\Gamma=0.5t$ ($g=0.5$). We discover that other spin components $s_{ix}$ and $s_{iy}$ are very small. (\textbf{d}) HF site spin values  at half-filling are shown.  Here, $U=t$, $n_{imp}=0.1$, and $\Gamma=0.5t$ ($g=0.16$ is smaller compared with the value used in (\textbf{c})).}
\label{dmrg-res}
\end{center}
\end{figure}

To model the graphene zigzag nanoribbons, we apply the Hubbard model with the nearest neighbor hopping and a diagonal disorder $V_i$.
\begin{eqnarray}
 H=-t\sum_{\langle ij \rangle,\sigma} c^{\dag}_{i,\sigma}c_{j,\sigma} +\sum_{i,\sigma} V_i c_{i,\sigma}^{\dag}c_{i,\sigma}
 +U\sum_i n_{i,\uparrow} n_{i,\downarrow},
\label{Hubbard}
\end{eqnarray}
where $i=(x,y)$ denotes the site indices (see Fig. \ref{dmrg-res}a), $c^{\dagger}_{i,\sigma}$/$c_{i,\sigma}$ are the creation/destruction operators at site $i$, $t$ is the nearest neighbor hopping parameter and  $U$ is the on-site repulsion. The ratio of the numbers of impurities and carbon atoms is given by $n_{imp}=N_I/N_s$. The values of the disorder strength $V_i$ at $N_I$ impurity sites  are  uniformly distributed in the interval 
$ [-\Gamma,\Gamma]$ (the sum in the second term of $H$ is only over impurity sites). The dimensionless coupling constant of the problem is the ratio of the disorder strength and on-site repulsion $g= \frac{\Gamma\sqrt{n_{imp}}}{U}$. The doping concentration is defined as $\delta N/N_s$, where $\delta N$ is the total number of added electrons. The mean field version of this Hamiltonian for a doped ribbon is given  in method, see Eq. (\ref{MFhspin}).
The HF results of undoped zigzag ribbons show that fractionalization occurs  independent of the disorder potential range, density, and strength. Note that disorder is a singular perturbation\cite{Jeong11,eeyang11}.

In graphene systems mean field approximations are widely used because they give accurate results\cite{Stau}.   \textcolor{black}{ However, there are several nearly degenerate HF ground states in graphene nanoribbons that can be generated using different HF initial states,and one does not know which of these states is close to the true ground state because quantum fluctuations are missing in the HF approximation. In this paper, we conducted the DMRG to determine which HF initial state generates the  HF state that is close to the true ground state. We will concentrate on two types of HF initial states in this section. The first, labeled AF, is generated from an antiferromagnetic initial state, while the second, labeled PM, is generated from a paramagnetic initial state with a small spin-splitting.}
The DMRG found that the \textcolor{black}{undoped} ground state at clean limit exhibits the N\'eel magnetic ordering, where spins at two zigzag edges align antiparallel to each other \textcolor{black}{(these results agree with those results obtained using the AF initial state)}. Nonetheless, the addition of enough extra electrons results in an edge spin density wave. The corresponding DMRG results, presented in Supplementary material,    \textcolor{black}{ agree with those results obtained using  the PM initial state, see Fig. \ref{dmrg-res}b.}
The results were then tested at half-filling for an undoped disordered interacting zigzag nanoribbon. The PM initial state produces a state with fractional charges. Site spin values computed from this ground state agree well qualitatively with those of the DMRG approach, as shown in \textcolor{black}{Figs.~\ref{dmrg-res}c,d.} We will show below 
that DMRG results for doped interacting disordered zigzag ribbons also support the results obtained from  the PM initial state. \textcolor{black}{The methods of both DMRG and HF approximation are explained in detail in Supplementary material.}

\subsection*{New Anyon Phase and TDOS of Low Doping Region}
\begin{figure}[htpb]
\begin{center}
\includegraphics[width=0.5\textwidth]{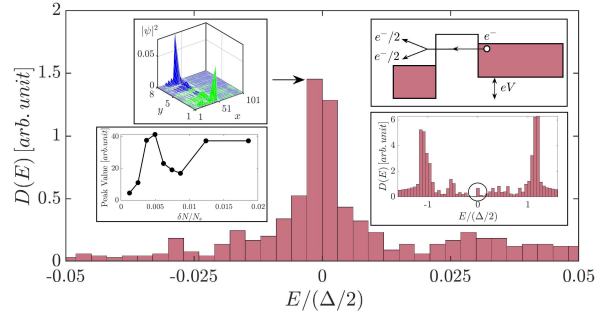}
\caption{\normalfont The HF results for $N_e=N_s+3$ ($\delta N/N_s=0.0037$), $\Gamma=0.01 t $, $L_x=101$, $L_y=8$, $n_{imp}=0.1$, and $U=t$ $(g=0.0032)$. DOS of a slightly doped ribbon away from half-filling is shown \textcolor{black}{in the weak disorder regime}. Sharp peak is present inside the soft gap at the midgap energy $E=0$  (the magnitude of this peak is rather {\it small} in comparison to the peaks at $E=\pm\Delta/2$ shown in Fig. \ref{degGap}). Since there are excess electrons, the Fermi energy $E_F/(\Delta/2)=0.14 $ is {\it above} the mid gap energy. \textcolor{black}{The DOS of $L=300$ in a larger  energy interval $E<\Delta$  is shown in the lower right inset.}
A charge fractionalized HF eigenstate is shown in the upper left inset. Note that energy is measured in units of $\Delta/2$. The number of disorder realization is $N_D\sim 400$. A tunneling electron is fractionalized in the upper right inset. The DOS is determined by measuring the differential I-V.  \textcolor{black}{Lower left  inset displays  the dependence of the midgap peak on doping concentration for $L=100$.}}
\label{SolPhas}
\end{center}
\end{figure}

\begin{figure}[!hbpt]
\begin{center}
\includegraphics[width=0.5\textwidth]{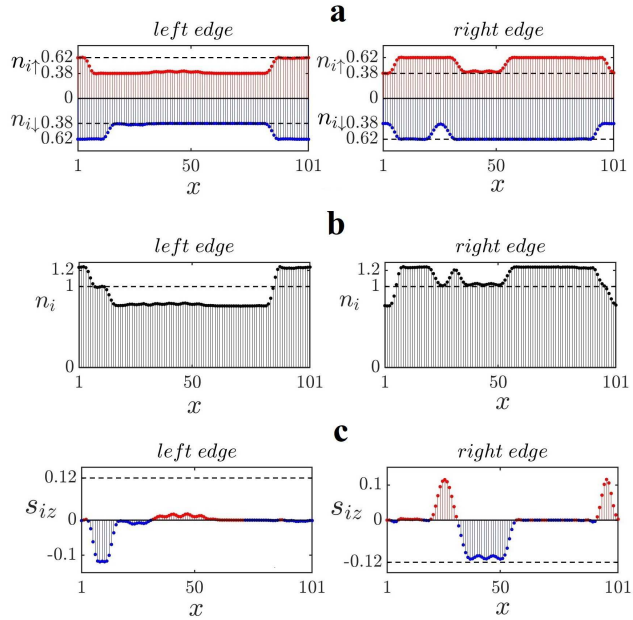}
\caption{\normalfont The HF results for $N_e=N_s+3$ ($\delta N/N_s=0.0037$), $\Gamma=0.01 t $, $L_x=101$, $L_y=8$, $n_{imp}=0.1$, and $U=t$ $(g=0.0032)$. Their disorder-free values are represented by dashed lines. (\textbf{a}) A disorder realization of zigzag edge site occupation numbers $n_{i\sigma}$ for a doped ribbon. (\textbf{b}) Total site occupation numbers $n_i$ are shown. Some charges are transferred between the zigzag edges on the left and right. (\textbf{c}) Site spins $s_{iz}$ are plotted. Their disorder-free values are represented by dotted lines.}
\label{SolPhas3}
\end{center}
\end{figure}

Using the PM initial state, we investigated the shape of the TDOS as a function of doping concentration \textcolor{black}{(all the HF results below are generated by using this HF initial state).}  As shown in Fig. \ref{SolPhas}, adding a few extra electrons to the half-filled ribbon results in a sharp peak near the midgap energy $E=0$ inside an exponentially decaying small soft gap. The peak's physical origin is as follows: A tunneling electron enters into a soliton state and divides into two fractionally charged quasiparticles
\begin{equation}
e^-\rightarrow e^-/2+e^-/2.
\end{equation}
(Ref.\cite{Girvin} gives a good account of this process). A soliton state is described by a non-local wave function, as shown in the upper left inset of Fig. \ref{SolPhas}.
The width of the central peak is $\sim 0.02\Delta/2$.
\textcolor{black}{In the low doping limit, when an entering electron has a non-zero energy $E\neq 0$, it has a significant chance not to split into $e^-/2$ charges because fractionalization is only approximate at non-zero energies\cite{eeyang11}. The lower left inset of Fig. \ref{SolPhas} shows the highly non-linear dependence of the peak value at E=0 on doping concentration $\delta N/N_s$. The zero energy peak in the DOS disappears for $\delta N/N_s>0.005$ (the shape of this DOS will be shown below). Such non-linear behavior is unusual and provides compelling evidence for fractional charges. The shape of the TDOS at the low doping limit differs significantly from that of the half-filled undoped state} (there are also two side peaks, one on each side of the sharp peak at the midgap energy. These peaks are not shown in Fig. \ref{SolPhas} because their energies  lie outside the energy range $|E|>0.05\Delta/2$). Edge site occupation numbers and site spins are displayed in Fig. \ref{SolPhas3}. These findings show modulated ferromagnetic edges.

\pagebreak

\begin{figure}[htpb]
\begin{center}
\includegraphics[width=0.7\textwidth]{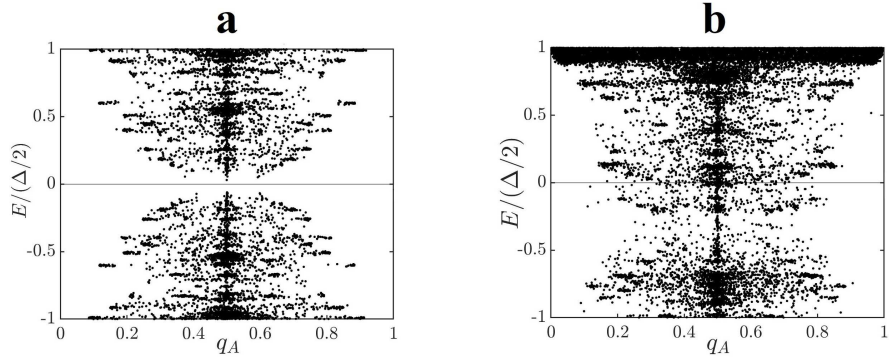}
\caption{\normalfont $q_A$ values of the HF eigenstates of a ribbon are plotted for (\textbf{a}) $N_e=N_s$  and (\textbf{b}) $N_e=N_s+3$ ($\delta N/N_s=0.0037$). A gap state electron with $q_A=1/2$ is fractionalized. Here $L_x=101$, $L_y=8$, $N_D=400$, $n_{imp}=0.1$, $U=t$, $\Gamma=0.01t$, and $N_e=N_s+3$ $(g=0.0032)$. In case of (\textbf{b}), the spectrum does not have particle-hole symmetry. }
\label{SolPhas1}
\end{center}
\end{figure}

 Many of the added electrons have $q_A$ values $\sim 1/2$,  which can be seen by comparing Figs. \ref{SolPhas1}a,b (the quantity $q_A$ gives the total probability to find an electron with energy $E$ on A-type carbon atoms. Charge fractionalization occurs when $q_A=1/2$). As a result, the extra electrons enter soliton states with well-defined fractional charges. Thus, our results for low doped ribbons indicate that doping does not destroy anyons. The average energy cost to create\cite{Girvin117} an $e^-/2$ fractional charge from the undoped ground state with an exponentially small gap is $\Delta_s/2$, which corresponds to the midgap energy $E=0$ in the excitation spectrum. This effect is similar to the formation of polyacetylene soliton midgap states\cite{yang1}. An undoped zigzag ribbon, unlike the chiral edges of Laughlin fractional quantum Hall states, lacks significant gapless edge excitations.

\begin{figure}[!hbpt]
\begin{center}
\includegraphics[width=0.3\textwidth]{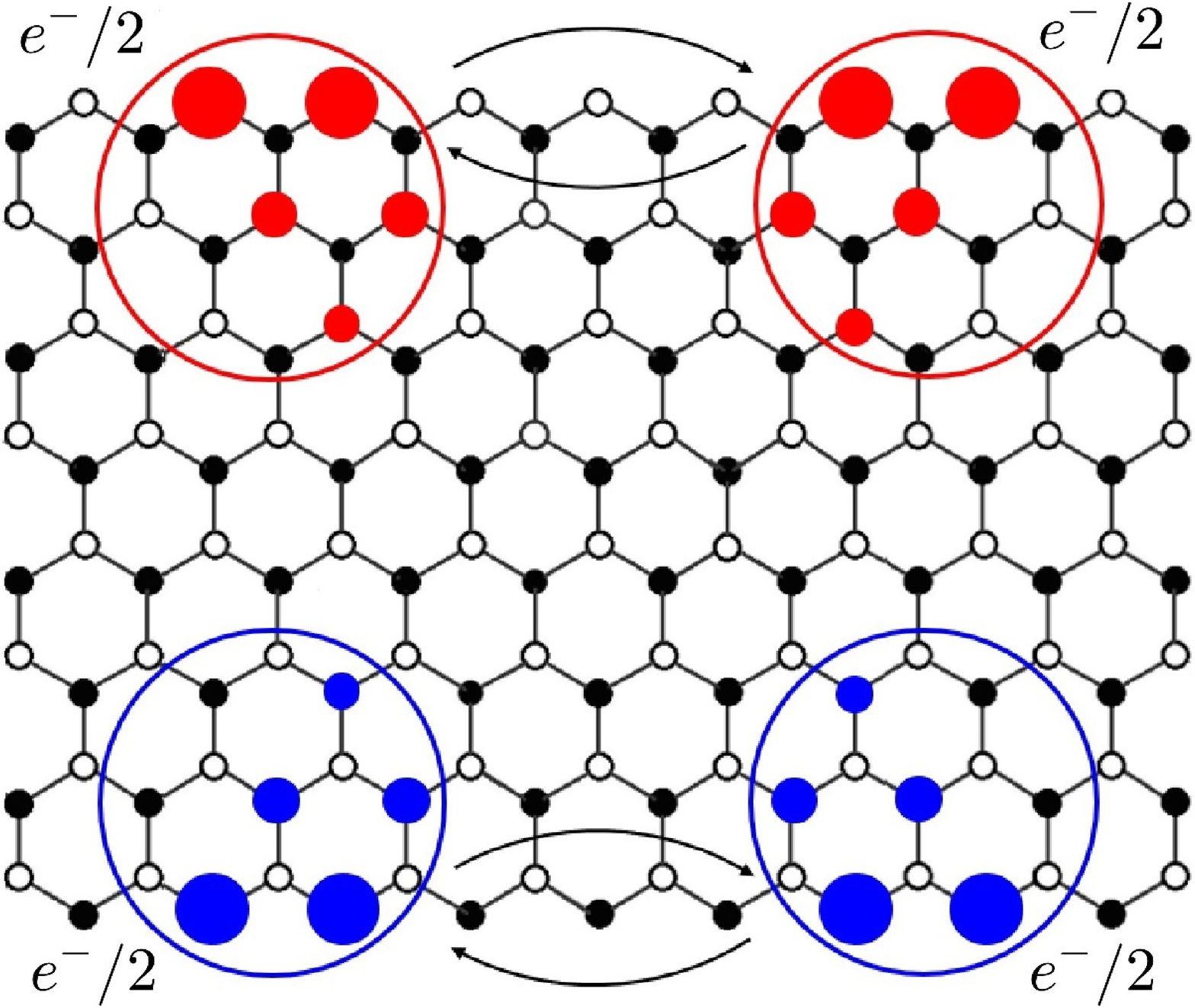}
\caption{\normalfont  The HF results for Two HF mixed chiral states are shown. Exchanges of two $e^-/2$ charges on the left zigzag edge and two others on the right zigzag edge are displayed.}
\label{semi}
\end{center}
\end{figure}

 Following  Ref.\cite{Wilczek03}  let us argue that an $e^-/2$ fractional charge of a disordered ribbon is an anyon.
Consider two single-particle HF mixed chiral states that display $e^-/2$ fractional charges, as shown in Fig. \ref{semi}. If we exchange these two electrons, the total many-body wave function of $N$ electrons acquires a statistical phase of $e^{i\pi}=-1$. This exchange is also equivalent to exchanging two $e^-/2$ charges on the left zigzag edge and two others on the right zigzag edge. Thus, we expect that each of these exchanges generates the statistical phase of $e^{i\pi/2}$ to yield the final phase of $e^{i\pi}=-1$. An anyon with the statistical phase $e^{i\pi/2}$ is called \textcolor{black}{a semion\cite{Kalme,Canri}.}
The presence of semions is consistent with the presence of anyons in TO phases.
It is also consistent with 
the shape of the entanglement spectrum.
 It is believed that the ground state entanglement spectrum of a TO phase resembles the corresponding edge spectrum of the system\cite{Haldane191,Van,fid}.
 The shape of the entanglement spectrum is computed and found to be similar to the DOS of the edge states (see Supplementary material).  The entanglement spectrum of undoped interacting disordered zigzag ribbons differs from that of zigzag ribbons in the disorder-free SPT phase\cite{Kim11}.


\begin{figure}[!hbpt]
\begin{center}
\includegraphics[width=0.5\textwidth]{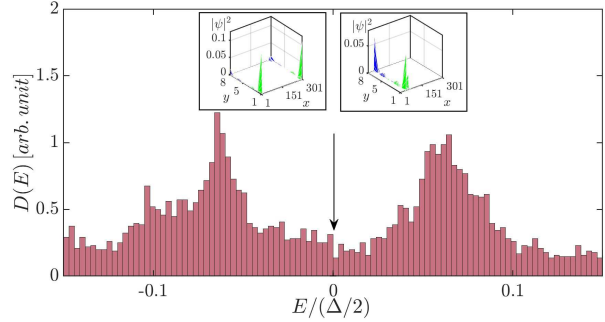}
\caption{\normalfont  The HF results for $N_e=N_s+20$ ($\delta N/N_s=0.0083$), $\Gamma=0.06 t$, $L_x=301$, $L_y=8$, $n_{imp}=0.1$, and $U=t$ ($g=0.019$). Two side peaks on the DOS. In the limit of large ribbon length or, equivalently, in the limit of zero doping, the profile of these two peaks becomes symmetric. Since there are excess electrons, the Fermi energy $E_F/(\Delta/2)=0.46$. The number of disorder realization is $N_D\sim 200$.}
\label{SolPhas0}
\end{center}
\end{figure}

\begin{figure}[!hbpt]
\begin{center}
\includegraphics[width=0.4\textwidth]{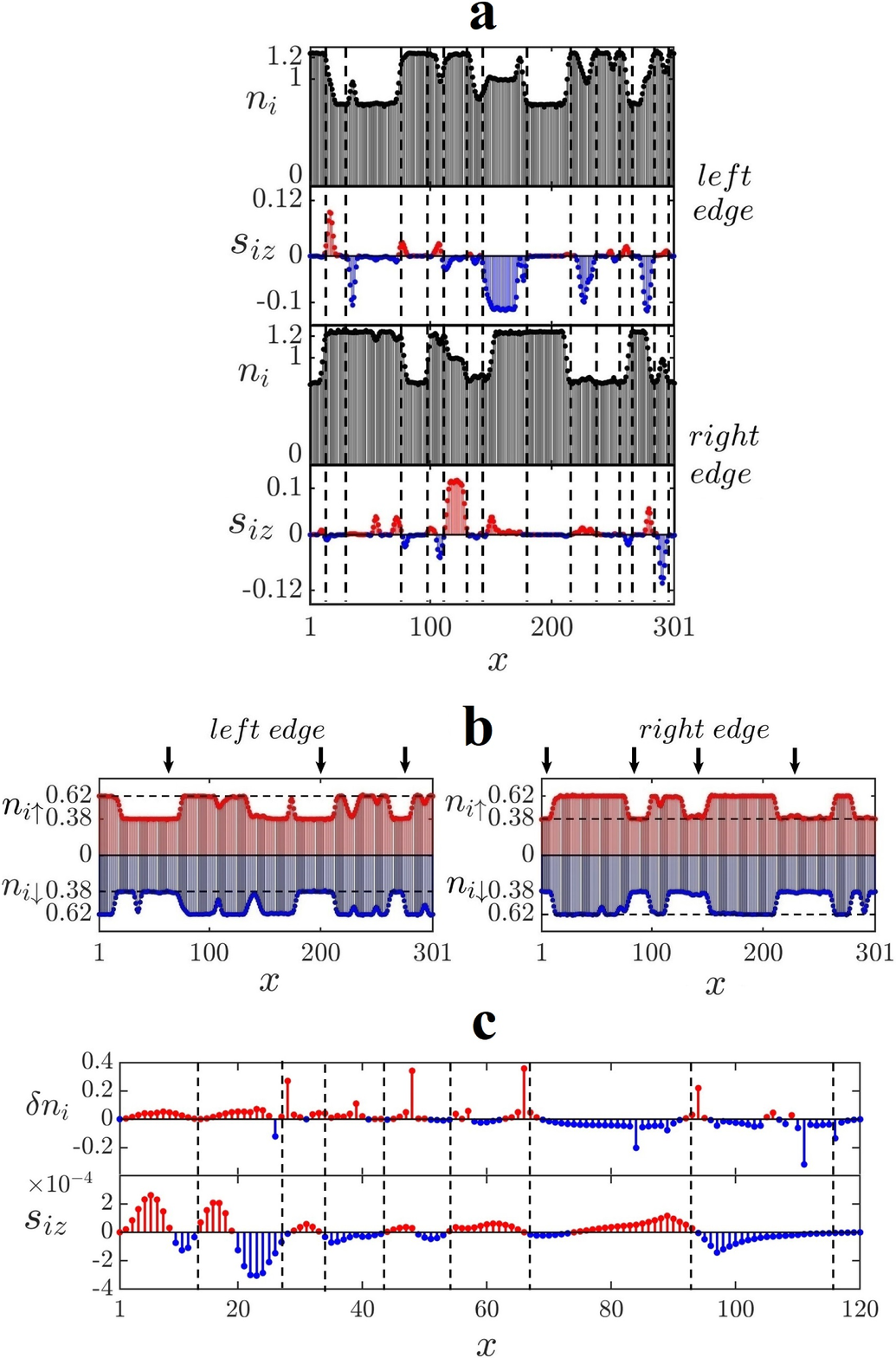}
\caption{\normalfont The HF results for $N_e=N_s+20$ ($\delta N/N_s=0.0083$), $\Gamma=0.06 t$, $L_x=301$, $L_y=8$, $n_{imp}=0.1$, and $U=t$ ($g=0.019$):   (\textbf{a}) Vertical lines indicate sites where $n_{i}$ or $s_{iz}$ abruptly change. On the zigzag edges, there are numerous localized magnetic moments. (\textbf{b}) Site occupation numbers $n_{i\sigma}$ of a disordered ribbon. The arrows point to locations where spin-charge separation occurs. (\textbf{c}) DMRG results for $\delta n_i \equiv n_i - n_i^{\rm clean}$ plotted as a function of $x$ and $y$. Here, $n_i^{\rm clean}$ is the site occupation number for $\Gamma=0$. Site spins $s_{iz}$ are also shown. The parameters are as follows:
$N_e=N_s+12$ ($\delta N/N_s=0.025$), $\Gamma=t$, $L_x=120$, $L_y=4$, $n_{imp}=0.2$, and $U=t$. Note that the length of this ribbon is considerably shorter than the one used in (\textbf{a}) and (\textbf{b}). These are the results for a more strongly disordered ribbon with $g=0.45$, and the overall magnetization is significantly reduced.  }
\label{SolPhas2}
\end{center}
\end{figure}

As more electrons are added, the sharp peak at the midgap energy  {\it disappears} in the DOS, but the two side peaks near $E\sim \pm 0.05\Delta/2 $ {\it persist}, as shown in Fig. \ref{SolPhas0}.
Simultaneously, the edge occupation number profile becomes highly nonuniform, as shown in Fig. \ref{SolPhas2}a.  These edge occupation numbers and site spin profiles appear to be quite different from those with fewer electrons, see Figs.~\ref{SolPhas3}a-c.
The nature of the disordered ground state of the doped system is as follows. To begin, it is important to note that the ground state of doped disorder-free zigzag ribbons differs from that of undoped ribbons, which have ferromagnetic edges that are antiferromagnetically coupled.  In sufficiently doped disorder-free ribbons both the DMRG (see Supplementary material) and HF display {\it spin density} type periodic modulations on the zigzag edges, see Fig. \ref{dmrg-res}b (the opposite edges are still antiferromagnetically coupled). The ground state changes again in the presence of disorder, and the periodic spin density is destroyed (a disorder potential is a {\it singular} perturbation\cite{eeyang11}). As the vertical dashed lines in Fig. \ref{SolPhas2}a show many HF values of $s_{iz}$ of each zigzag edge change
sign at sites where $n_i$ abruptly changes (see Fig. \ref{SolPhas2}b).
Similar behavior is also observed in the DMRG calculation, as shown in Fig. \ref{SolPhas2}c. 
Moreover, sites $i$, where the values of $n_{i\sigma}$ and $s_{iz}$ abruptly change, have almost identical values for the $x$-coordinate on the left and right zigzag edges. This effect is a consequence of the {\it nonlocal} correlation between the left and right zigzag edges. This correlation between opposite zigzag edges strongly suggests that the formation of non-local soliton states is responsible for the drastic reorganization of the ground state, as well as the zigzag edge modifications.
Furthermore, we find {\it local magnetic moments} with non-zero values of $s_{iz}$ that is extended over several sites.
These objects
 proliferate in comparison to the case of undoped ribbons.  There are also objects extended over several sites with rather small values of
$s_i=\frac{1}{2}(n_{i\uparrow}-n_{i\downarrow})\approx 0$, see Fig. \ref{SolPhas2}b. In such an object, {\it spin-charge separation} would take place. The following procedure is used to create these objects. An $e^-/2$ fractional charge moves along the zigzag edges from left to right, while another fractional charge with the opposite spin moves in the opposite direction (see Ref.\cite{eeyang11} for a detailed explanation).

The resulting ground state displays a highly distorted edge spin density.
This phase is characterized by localized edge magnetic moments, spin-charge separation, and correlation between the left and right zigzag edges, a {\it disordered anyon phase} of zigzag nanoribbons.  \textcolor{black}{Charge fractionalization is not exact in this phase since some of the nearly zero energy states are not fractionalized, see the left inset in Fig. \ref{SolPhas0}. This is in contrast to the results of slightly doped and undoped disordered ribbons, where the fractional charge of zero energy  states is well-defined (see Fig. \ref{SolPhas} of the current manuscript and Fig.9 in Ref.\cite{eeyang11}, respectively).}
When doping concentration is increased further ($\delta N/N_s\sim 0.04$), the distorted edge spin density wave   and   charge fractionalization almost disappear. We also discover that the HF gap states are no longer localized along  the ribbon direction.   \textcolor{black}{ These findings imply that as doping concentration increases from zero, a topological phase transition with a significant crossover region occurs.}


\newpage
\section*{Discussion}
Quantum fluctuations beyond the HF approximation do not mitigate charge fractionalization in a ribbon at low doping concentration, according to our findings. Despite the presence of disorder, as doping increases, the edge magnetic ordering weakens and charge fractionalization disappears. Furthermore, we discovered that the low doped state is a new disordered anyon phase with highly distorted edge charge and spin modulations, as well as localized magnetic moments with non-local correlations between the left and right zigzag edges. Anyons play a key role in the formation of this new phase.
 As a result of spin-charge separation, our findings suggest that doped zigzag nanoribbons may exhibit new magnetic and low temperature transport properties: the conductivity may display a usual behavior while the spin susceptibility may be rather small, as was observed in polyacetylene\cite{Chung15}.
Furthermore, we demonstrated that the TDOS profile is significantly affected by doping concentration. The measurement of the differential  I-V curve may reveal this effect and may provide a strong test for the presence of $e^-/2$ fractional charges. \textcolor{black}{Ribbons with width less than $100$~\AA\,  are
well-suited for the observation of these fractional charges as the  antiferromagnetic phase  is more stable than  the ferromagnetic phase\cite{Pisa1}.}

The following additional investigations may be interesting to pursue. A worthwhile but challenging task is to compute the anyon statistical phase using a microscopic approach. Recently such an adiabatic DMRG simulation, utilizing the quantized Hall response,  was successfully conducted\cite{Zhu} for the non-Abelian Moore-Read state on a Haldane honeycomb lattice model.
A similar DMRG calculation in a Mott-Anderson insulator of disordered zigzag nanoribbon with Abelian quasiparticles, where electron localization is critical, is not clear.
Another method is to compute the statistical Berry phase of Abelian quasiparticles using a trial wave function\cite{Arovas}. But a good trial wave function is not yet available for disordered zigzag ribbons.

\textcolor{black}{In the limit of small doping and in weak disorder regime, ribbons  with a sharp midgap peak in the DOS have a universal value of the  TEE\cite{Kim11}. When doping is high enough, the midgap peak disappears, as does the exact fractionalization of zero energy states. When this occurs, we can expect non-universal TEE values. It may be worthwhile }to  probe the topological phase transition as  doping concentration increases. The following issues must be addressed: \textcolor{black}{Does} the TEE of doped disordered ribbons decay to zero as a function of doping concentration, and does the transition exhibit non-universal dependence on physical parameters\cite{Kim11}?  This investigation  may require very high cpu resources to compute accurately small values of the TEE\cite{Kim11}. This type of calculation could provide more information about the phase transition from modulated ferromagnetic edges at zero doping to distorted spin-wave edges at finite doping

The following experiments would also be fascinating. Investigation of tunneling between zigzag edges, as seen in fractional quantum Hall bar systems \cite {Kang}, may be fruitful.
 Scanning tunneling microscopy can reveal the presence of fractional charges by measuring the electron density on the zigzag edges\cite{Andrei}. Finally, it would be interesting to look into the new disordered anyon phase in other antiferromagnetic zigzag nanoribbon systems, e.g., silicene and boron nitride nanoribbons \cite{Yao,Bar}. Chiral gauge theory can be used to describe $e^-/2$ fractional charges\cite{ChiralTH}.  It would be fascinating to look into the new anyon phase using random chiral gauge fields.
\section*{Methods}
\subsection*{HF Approximation}
To model graphene zigzag nanoribbons, the mean field Hubbard model is commonly used \begin{eqnarray}
H_{MF}=-t\sum_{\langle ij\rangle,\sigma} c^{\dag}_{i,\sigma}c_{j,\sigma} +\sum_{i,\sigma} V_ic_{i,\sigma}^{\dag}c_{i,\sigma}
+U\sum_i[ n_{i,\uparrow}\langle n_{i,\downarrow}\rangle +n_{i,\downarrow}\langle n_{i,\uparrow}\rangle-\langle n_{i,\downarrow}\rangle\langle n_{i,\uparrow}\rangle]
+ \sum_i [s_{ix} \langle h_{ix}\rangle+s_{iy} \langle h_{iy}\rangle]
\label{MFhspin}
\end{eqnarray}
where $ \langle h_{ix}\rangle=-2U\langle s_{ix}\rangle$ and $ \langle h_{iy}\rangle=-2U\langle s_{iy}\rangle$ are the self-consistent ``magnetic fields."
The last term of Eq. (\ref{MFhspin}) describes spin flips and is present but only separately from half-filling. This term mixes spin-up and spin-down. Note that the band structure no longer has particle-hole symmetry when away from half-filling.\\

\subsection*{DMRG}
We apply the DMRG\cite{White1992, Rommer,Schollwock2011} to obtain the ground state of the model Eq.~(\ref{Hubbard}) in the MPS representation.
Furthermore, we illustrate the geometry of the MPS for the graphene zigzag nanoribbon of the size $(L_x \times L_y)$ (see Supplementary material).
For a quasi-one-dimensional system, the complexity of the DMRG scales exponentially in the width of the system\,($L_y$), whereas it scales polynomially in the length\,($L_x$) of the system. Therefore, our MPS setup allows us to consider the graphene strip with long zigzag edges, and we focus on the system with $(L_x,L_y) = (120,4)$, which is far beyond the reach of exact diagonalization in the present calculations. The precision of the DMRG can be controlled by the number of basis states kept or the maximum bond dimension of the MPS ($\chi_{\rm max}$), and we use up to $\chi_{\rm max} = 1600$ to achieve the typical error of the total energy lower than $10^{-6}$ (For a short introduction to MPS, see Supplementary material). To fix a gauge redundancy of MPS specially, i.e., the canonical form, the DMRG optimizes each tensor considering the global information of the wave function, which makes the algorithm extremely stable and reliable\cite{Schollwock2011}. Nonetheless, the DMRG can become trapped in a local minimum, particularly for models with a quasi-one-dimensional lattice. To avoid local minima, we apply the noise perturbation\cite{White05} with the two-site algorithm\cite{Schollwock2011} at each optimization step.
Moreover, we exploit the $U(1)$ symmetry of the model such that the DMRG preserves the total number of electrons, e.g., $\sum_{i,\sigma} c^\dag_{i,\sigma} c_{i,\sigma} = N_s/2$, thereby improving greatly its convergence speed and accuracy\cite{Schollwock2011}.

\section*{Data availability}
On reasonable request, the corresponding author will provide all relevant data in this paper.

\section*{Code availability}
On reasonable request, the corresponding author will provide all numerical codes in this paper.


\section*{Acknowledgments}

S.R.E.Y. was supported by the Basic Science Research Program through the National Research Foundation of Korea (NRF), funded by the Ministry of Science and ICT (MSIT) NRF-2021R1F1A1047759.
The DMRG simulations were partially performed on the Cedar cluster, hosted by WestGrid and Compute Canada. H.Y.L. was supported by the National Research Foundation of Korea under the grant numbers NRF-2020R1I1A3074769 and the BK21 FOUR (Fostering Outstanding Universities for Research).

\section*{Author contributions}

Y.H.K and H.J.L. performed the HF calculations. H.Y.L. performed the DMRG computations. S.R.E.Y conceived the project and supervised the study. All authors contributed to the writing of the manuscript.

\section*{Competing interests}
The authors declare no competing interests.

\section*{Additional information}

The Supplementary material is available.
\end{document}


\title{Supplementary material: New disordered anyon phase of doped graphene zigzag nanoribbon}

\author{Young Heon Kim\orcidlink{0000-0003-0000-539X}}
\affiliation{Department of Physics, Korea University, Seoul 02855, Korea}
\author{Hye Jeong Lee\orcidlink{0000-0003-4535-9511}}
\affiliation{Department of Physics, Korea University, Seoul 02855, Korea}
\author{Hyun-Yong Lee\orcidlink{0000-0003-4774-5213}}
\affiliation{Department of Applied Physics, Graduate School, Korea University, Sejong 30019, Korea}
\affiliation{Division of Display and Semiconductor Physics, Korea University, Sejong 30019, Korea}
\affiliation{Interdisciplinary Program in E$\cdot$ICT-Culture-Sports Convergence, Korea University, Sejong 30019, Korea}
\author{S.-R. Eric Yang\orcidlink{0000-0003-3377-1859}}
\email{corresponding author: eyang812@gmail.com}
\affiliation{Department of Physics, Korea University, Seoul 02855, Korea}
\date{\today}

\maketitle

\section{Matrix Product States}

\begin{figure}[htpb]
\begin{center}
	\includegraphics[width=0.4\textwidth]{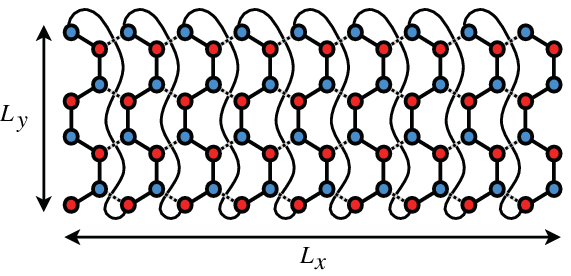}
\caption{Schematic figure of the snake pattern of the matrix product state defined on the zigzag nanoribbon lattice.}
\label{fig:mps_geometry}
\end{center}
\end{figure}

A large class of quantum many-body wavefunctions, $\Psi$, can be efficiently factorized into a product of tensors as follows:
\begin{align}
	|\psi\rangle &= \sum_{ \{ s_i \} } \Psi_{\cdots s_{i-1} s_i s_{i+1} \cdots} \,\, \big| \cdots s_{i-1} s_i s_{i+1} \cdots \big\rangle \nonumber\\
	&= \sum_{ \{ s_i \} } {\rm tTr}[ \cdots A_{i-1}^{s_{i-1}} A_i^{s_i} A_{i+1}^{s_{i+1}} \cdots ] \big|\cdots s_{i-1} s_i s_{i+1} \cdots \big\rangle,
\end{align}
where $A_{i}^{s_i}$ stands for a tensor at site $i$, $\cdots A_{i-1}^{s_{i-1}} A_i^{s_i} A_{i+1}^{s_{i+1}} \cdots$ indicates the product of tensors forming a network, depending on how tensors are connected, and ${\rm tTr[\cdots]}$ stands for the tensor trace or contraction of all connected indices in the network. Here the index $s_i$ denotes a local state at site $i$, e.g., $|s_i\rangle = |\uparrow\rangle, |\downarrow\rangle$ for a spin-half fermion system or $|s_i\rangle = |0\rangle, |\uparrow\rangle, |\downarrow\rangle, |\uparrow\downarrow\rangle$ for a spin-full fermion system. Particularly, the  so-called MPS, which is a chain-like product of rank-3 tensors, may represent quantum states for one-dimensional and quasi-one-dimensional systems accurately. Specifically, the many-body wave function in the MPS representation is written as follows:

\begin{align}
	\Psi_{\cdots s_{i-1} s_i s_{i+1} \cdots}	
 	= \sum_{\{l_i\}, \{r_i\}} \cdots \,\, \delta_{r_{i-2} l_{i-1}} \,\, [A_{i-1}^{s_{i-1}}]_{l_{i-1}, r_{i-1}} \,\, \delta_{r_{i-1} l_i} \,\, [A_i^{s_i}]_{l_i, r_i} \,\, \delta_{r_i l_{i+1}} \,\, [A_{i+1}^{s_{i+1}}]_{l_{i+1}, r_{i+1}} \,\,\delta_{r_{i+1} l_{i+2}} \,\, \cdots,
\end{align}

where $[A_i^{s_i}]_{l_i, r_i}$ is a rank-3 tensor with two virtual indices $l_i$ and $r_i$\,(say left and right, respectively), which are traced out, $s_i$ is the physical index, and $\delta_{ij}$ stands for the Kronecker delta or the identity matrix. Note that the Kronecker delta contracts the right and left indices of tensor at $i$ and $i+1$, respectively. Hence, the element of the wavefunction for a given set of $\{s_i\}$ is identical to the product of {\it matrices} $\{A_i^{s_i}\}$, i.e., $\Psi_{\cdots s_{i-1} s_i s_{i+1} } = \cdots A_{i-1}^{s_{i-1}} A_i^{s_{i}} A_{i+1}^{s_{i+1}} \cdots$. It is also convenient to introduce a graphical representation for the tensor and its network. A tensor is depicted by an object with open legs denoting its indices. For instance, the tensor $A_i^{s_i}$ can be illustrated as follows:
\begin{align}
	\includegraphics[width=0.2\textwidth]{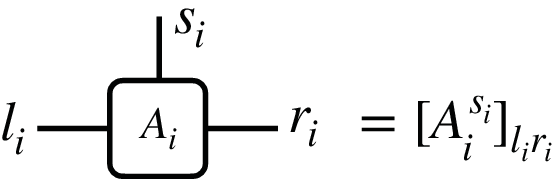},
\end{align}
where the vertical open leg $s_i$ denotes the physics index and the horizontal ones $l_i$ and $r_i$ stand for the virtual indices. Furthermore, a contraction of two indices, particularly one left and one right horizontal indices, occurs by connecting the legs\,(or indices) as shown below:
\begin{align}
	\includegraphics[width=0.4\textwidth]{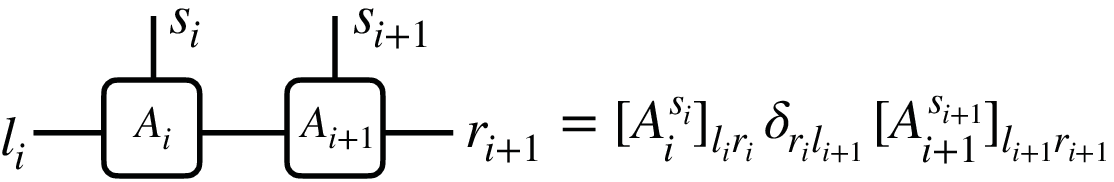}.
\end{align}
Thus, the total wavefunction in the MPS representation is illustrated as follows:
%
\begin{align}
	\includegraphics[width=0.48\textwidth]{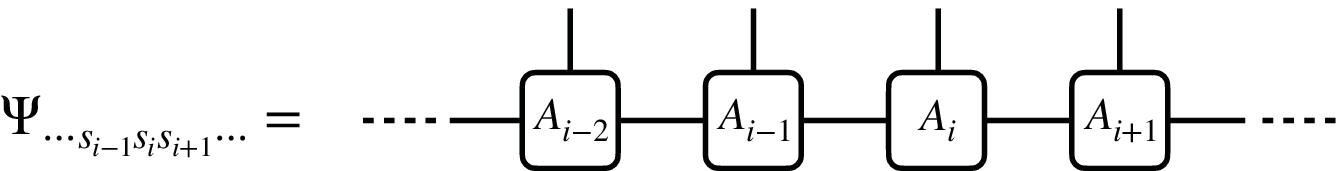}.
\end{align}
%
Here, the size of the matrix $A_i^{s_i}$, which is referred to as the bond dimension $\chi$, determines the expressibility of the MPS. In other words, the accuracy of the MPS can be systematically enhanced by increasing the value of $\chi$. We define the MPS on the zigzag nanoribbon in a so-called snake pattern, as illustrated in Fig.~\ref{fig:mps_geometry}. This definition allows us to consider the graphene strip with long zigzag edges.

Similarly, a large class of Hamiltonians can be represented exactly as a product of matrices or as the matrix product operator\,(MPO) as follows:
\begin{align}
	\hat{H} &= \sum_{\{s_i\},\{s_i'\}} H_{\cdots s_{i-1} s_i s_{i+1} \cdots}^{\cdots s_{i-1}' s_i' s_{i+1}' \cdots} \,\,\big|\cdots s_{i-1}' s_i' s_{i+1}' \cdots \big \rangle \big\langle \cdots s_{i-1} s_i s_{i+1} \cdots \big| \nonumber \\
	& = \sum_{\{s_i\},\{s_i'\}} {\rm tTr}[ \cdots W_{i\!-\!1,s_{i\!-\!1}}^{s_{i\!-\!1}'} W_{i,s_{i}}^{s_{i}'} W_{i\!+\!1,s_{i\!+\!1}}^{s_{i\!+\!1}'} \cdots ] \big|\cdots s_i' \cdots \big \rangle \big\langle \cdots s_i \cdots \big|,
\end{align}
where $W_{i,s_i}^{s_i'}$ is a rank-4 tensor with its graphical representation given as
\begin{align}
	\includegraphics[width=0.2\textwidth]{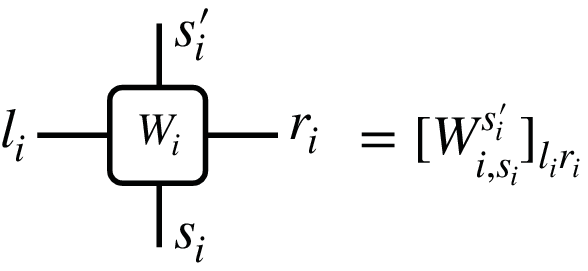}.
\end{align}
Then, the Hamiltonian is represented in the graphical representation as follows:
%
\begin{align}
	\includegraphics[width=0.48\textwidth]{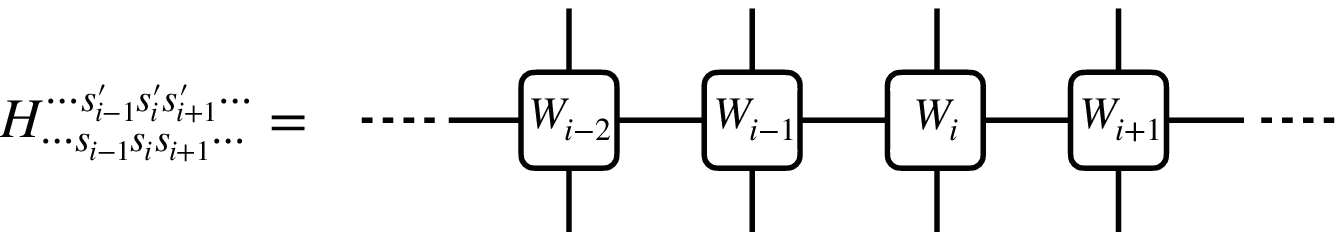}.
\end{align}
%
For example, the Hamiltonian of the one-dimensional Anderson--Hubbard model is produced with the tensor $W_{i,s}^{s'}$ as follows:
%

\begin{align}
	[W_{i,s}^{s'}]_{lr} = 	
	\begin{bmatrix}
		(I_i)_{ss'}	& 0 & 0 & 0 & 0 & 0 \\
		(c_{i\uparrow}^\dagger)_{ss'} 	& 0 & 0 & 0 & 0 & 0 \\
		(c_{i\downarrow}^\dagger)_{ss'} 	& 0 & 0 & 0 & 0 & 0 \\
		(c_{i\uparrow})_{ss'} 	& 0 & 0 & 0 & 0 & 0 \\
		(c_{i\downarrow})_{ss'} 	& 0 & 0 & 0 & 0 & 0 \\
		(U n_{i\uparrow} n_{i\downarrow} + v_i (n_{i\uparrow} + n_{i\downarrow}))_{ss'} & t (c_{i\uparrow})_{ss'} & t (c_{i\downarrow})_{ss'} & -t (c_{i\uparrow}^\dagger)_{ss'} & -t (c_{i\downarrow}^\dagger)_{ss'} & (I_i)_{ss'}
	\end{bmatrix}_{lr},
\end{align}

%
where $I_i$ is the trivial operator acting on site $i$, $c_i^{(\dagger)}$ is the annihilation\,(creation) operator, $v_i$ is the Anderson random potential, and $U$ is the Hubbard interaction. One can easily generalize the above $W-$tensor and MPO into the proper quantities for the model of the quasi-one-dimensional honeycomb lattice.

The total energy of a given wavefunction is given as
\begin{align}
	E = \frac{\langle \Psi | H | \Psi \rangle}{\langle \Psi | \Psi \rangle}
	= \frac{\sum_{\{s_i\},\{s_i'\}} (\Psi_{\cdots s_i' \cdots})^* H_{\cdots s_i \cdots}^{\cdots s_i' \cdots} (\Psi_{\cdots s_i \cdots})}
	{\sum_{\{s_i\}} (\Psi_{\cdots s_i \cdots})^* (\Psi_{\cdots s_i \cdots})},
\end{align}
and this total energy is graphically recast as
\begin{align}
	\includegraphics[width=0.4\textwidth]{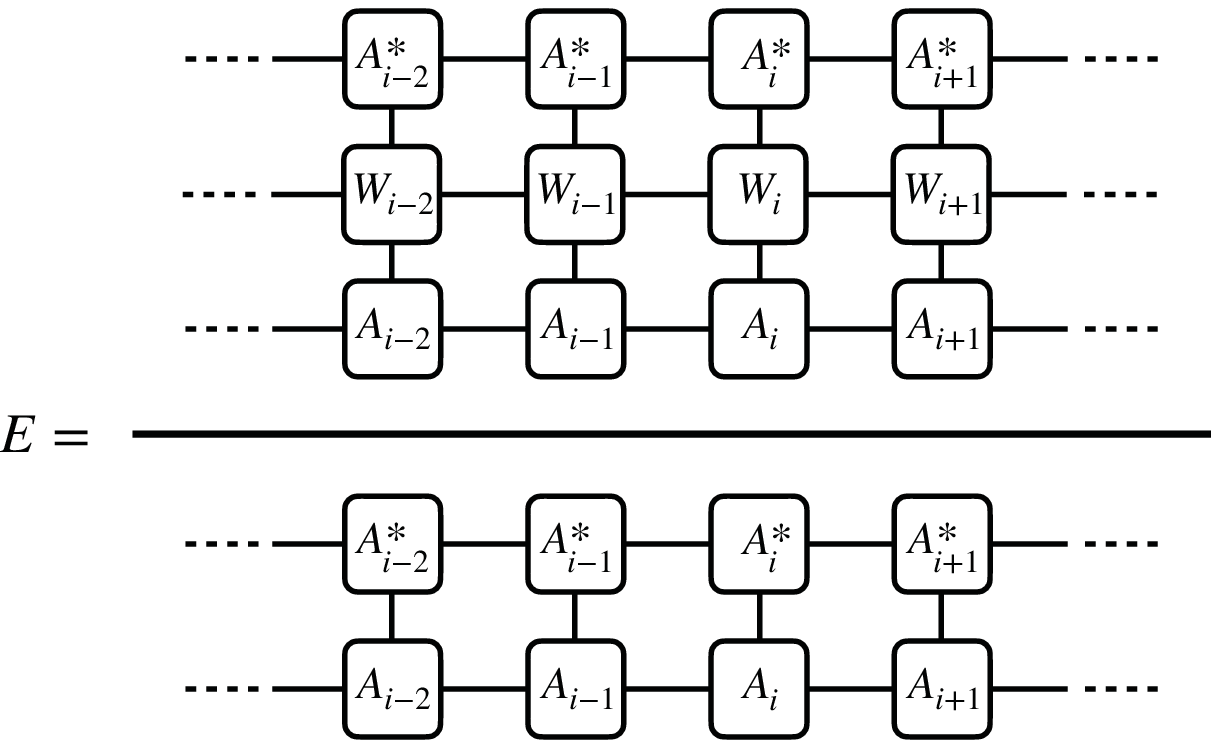}.
\end{align}
Thus, to measure the total energy $E$, one should contract two different tensor networks in the denominator and numerator in the above equation. The contraction of tensor networks can be efficiently conducted by choosing the order of the contraction properly~\cite{Schollwock2011}. With the above expression, one can apply a variational principle to optimize tensors $\{A_i\}$ by minimizing the energy, e.g.,
%
	$\partial E/\partial A_i^* = 0$,
%
Thereby leading to an eigenvalue problem to update the tensor $A_i$\,(see Ref.~\cite{Schollwock2011} for more details).

\subsection{Density Matrix Renormalization Group: Clean limit}

\begin {figure}[htpb]
\begin{center}
\includegraphics[width=0.49\textwidth]{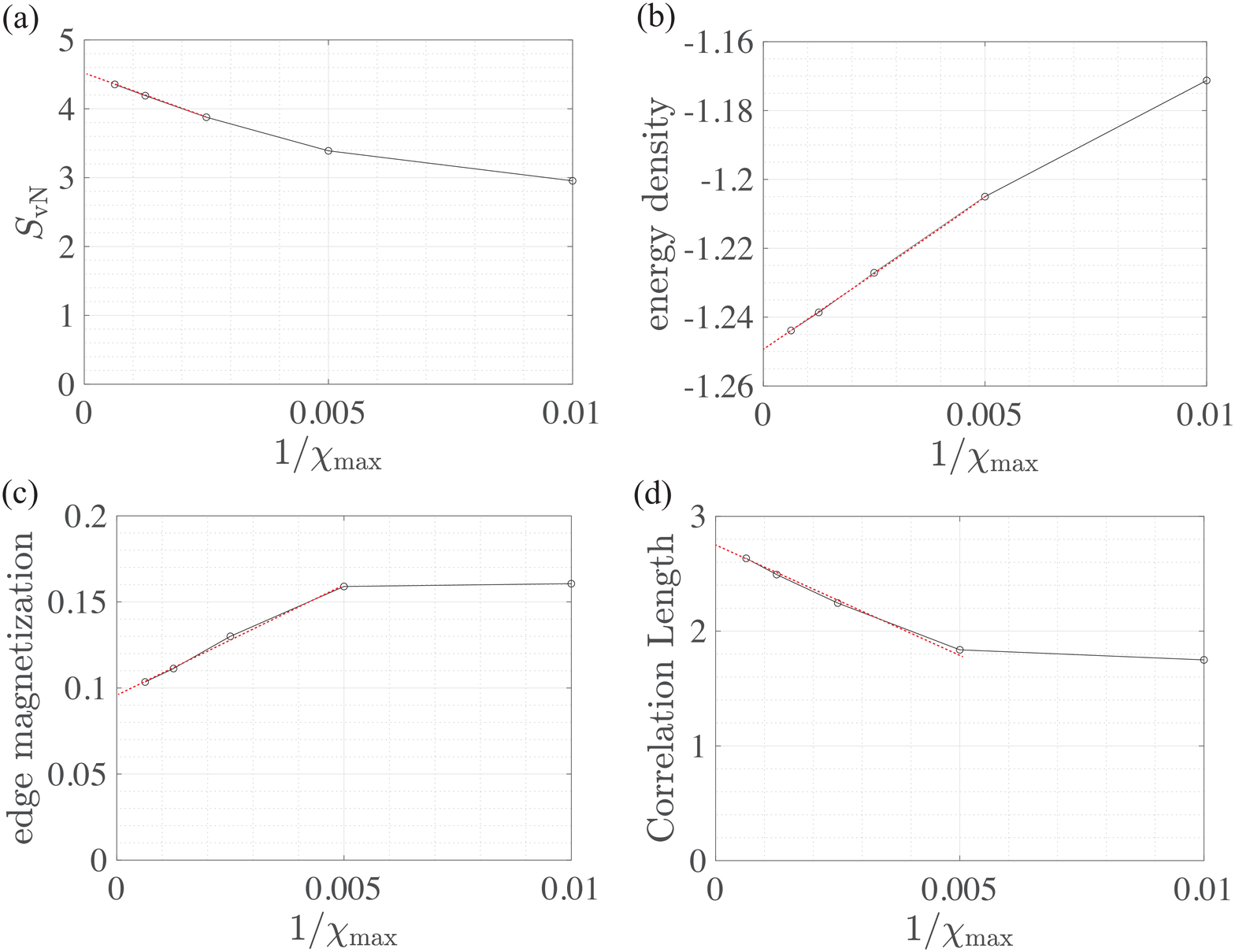}
\caption{The bond dimension scaling of entanglement entropy (a), energy density (b),  edge magnetization ($\hbar=1$) (c),  and  correlation length (d).  Here  $U/t=1$ is used.   The energy  unit is $t$ and the length unit is $\sqrt{3}a$, where  $a$ is the carbon-carbon distance.}
\label{fig:idmrg_scaling}
\end{center}
\end{figure}

As discussed in the previous section, the bond dimension of the MPS determines the amount of the MPS. The number of MPS should be carefully chosen to appropriately represent the ground state. Specifically, the MPS is optimal for representing quantum states that satisfy the area law of the entanglement, i.e., its entanglement entropy scales as the size of the boundary between two subsystems. By contrast, the entanglement entropy of a gapless ground state diverges logarithmically with the subsystem size. Thus, we first check whether the ground state of our Hamiltonian is gapped or gapless. Then, we determine how large the bond dimension should be to appropriately represent the ground state. To this end, we perform the infinite-size variant of the DMRG ~\cite{Schollwock2011} and see how the energy density, total magnetization, entanglement entropy scale and correlation length as a function of the maximum bond dimension $\chi_{\rm max}$. The result is presented in Fig.~\ref{fig:idmrg_scaling}. It is certain that the entanglement entropy does not diverge but converges to a finite value as $\chi_{\rm max} \rightarrow \infty$\,[see red dashed line in Fig.~\ref{fig:idmrg_scaling}(a)], which implies that the ground state is gapped or satisfies the area law of the entanglement. Also, based on the scaling result of the energy density and magnetization,\, as shown in Figs.~\ref{fig:idmrg_scaling}(a) and \ref{fig:idmrg_scaling}(b), we conclude that keeping $1600$ states, i.e., $\chi_{\rm max}=1600$, is enough to capture the essential physics of the ground state. Thus, we fix $\chi_{\rm max}$ to 1600 in all calculations. Although we did not present the results here, we however directly confirmed that keeping 1600 states provides convergent results in the finite system regardless of disorder.

\begin{figure}[htpb]
\begin{center}
\includegraphics[width=0.49\textwidth]{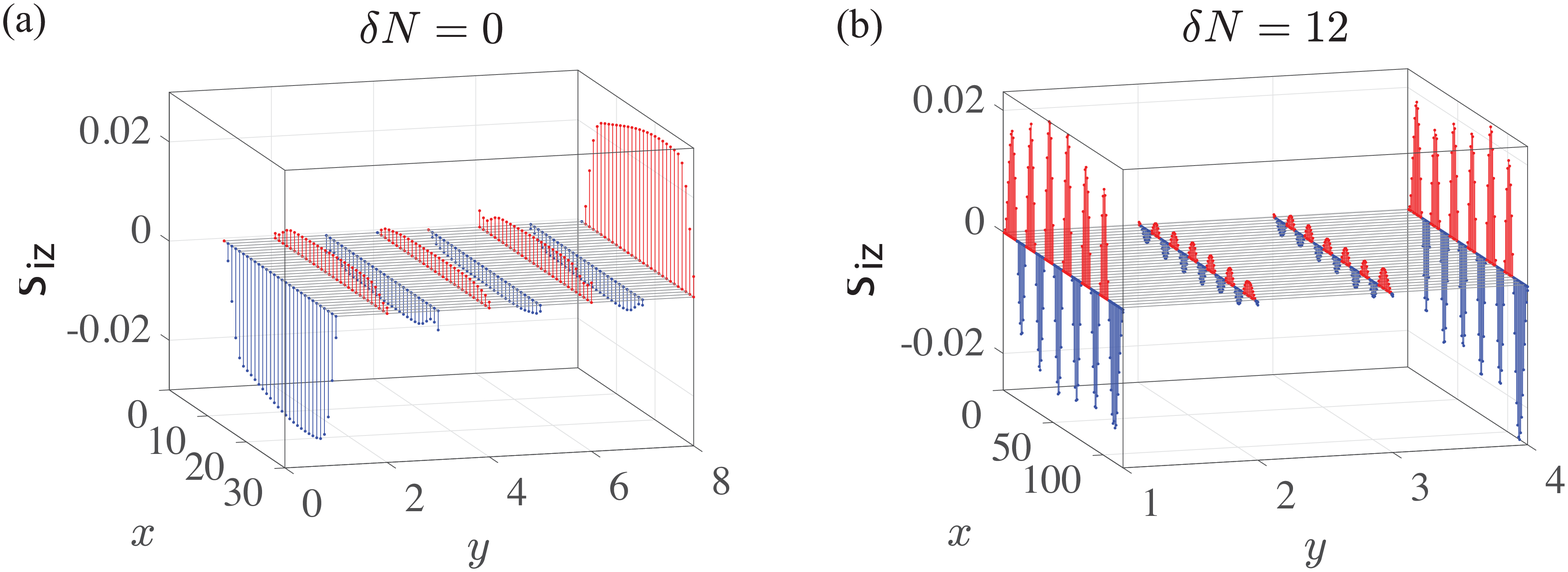}
\caption{Magnetization profile at (a) $(L_x,L_y, \delta N) = (30, 8, 0)$ and (b) $(L_x, L_y, \delta N) = (120, 4, 12)$ in the clean limit with $U/t=1$.}
\label{fig:dmrg_clean}
\end{center}
\end{figure}

As a benchmark result, we present in Fig.~\ref{fig:dmrg_clean} the DMRG results of magnetization profile in the clean system. The system is at half-filling\,($\delta N = 0$) with size $(L_x, L_y) = (30,8)$, as shown in Fig.~\ref{fig:dmrg_clean}(a). Expectedly, the ground state shows the N\'eel order where spins at each zigzag edge align antiparallel to each other. In Fig.~\ref{fig:dmrg_clean}(b), the system is slightly away from the half-filling, i.e., $\delta N = 12$ with size $(L_x, \, L_y) = (120,4)$. The doping introduces a spin-density wave. Hence, the magnetization oscillates along the ribbon direction.

%

\section{HF Approximation and Topological Entanglement Spectrum.}%

The ground state entanglement spectrum~\cite{Haldane191} is computed at half-filling using the HF approach. The  spectrum differs from that of disorder-free SPT zigzag ribbons: degenerate eigenvalues of the SPT phase are split and distributed similarly to that of the DOS of the edge states of a disordered TO zigzag ribbon.
The entanglement spectrum may be obtained from the reduced density matrix.
To compute the bulk entanglement spectrum, we choose the region~\cite{Bal} $A$ separate from the zigzag edges. We use the HF approximation; thus, the relevant reduced density matrix for a region $A$ can be written as~\cite{Peschel119}
\begin{eqnarray}
\rho_A=K e^{ -\tilde{h}}.
\end{eqnarray}

\begin{figure}[!hbpt]
\begin{center}
\includegraphics[width=0.3\textwidth]{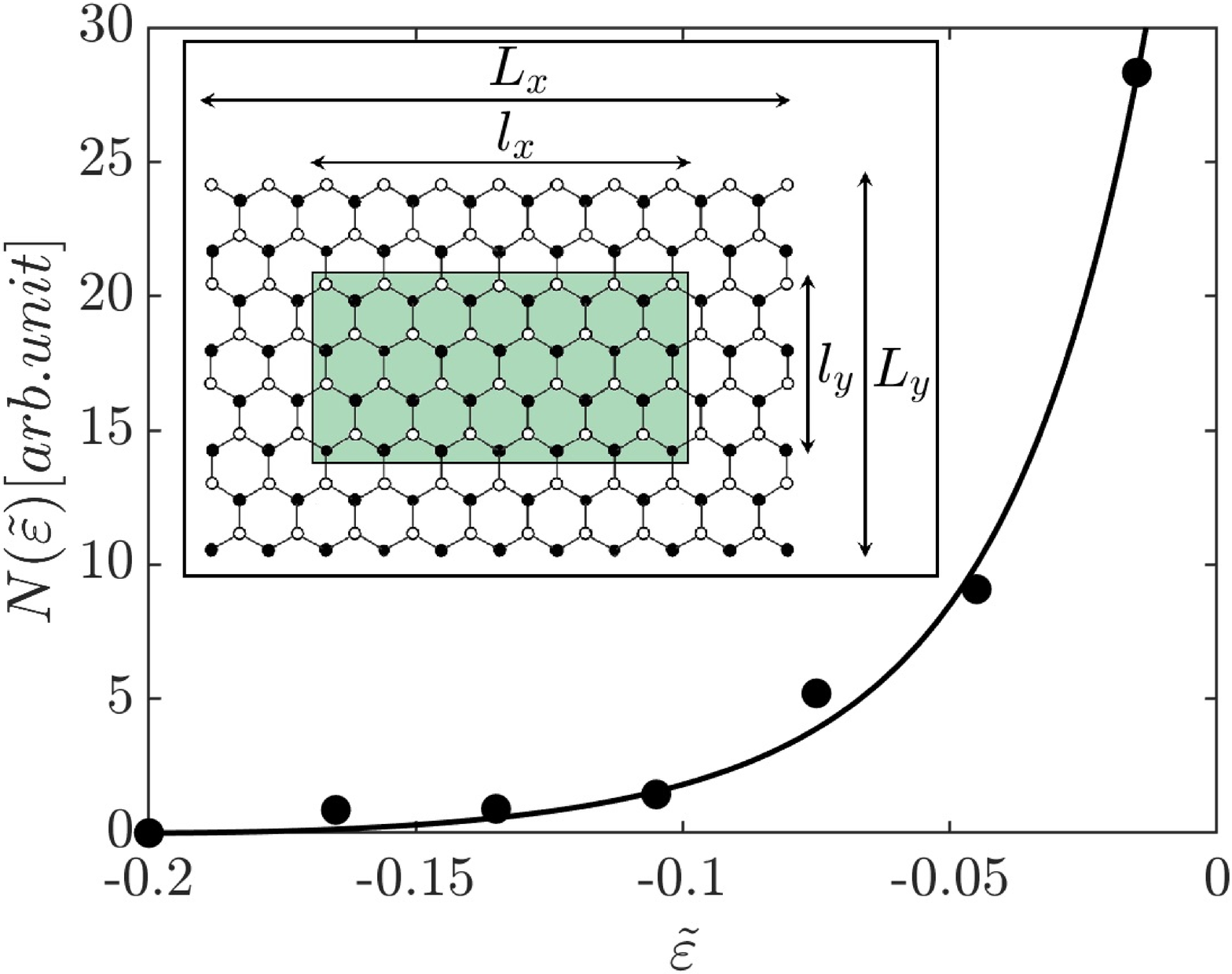}
\caption{Consider a rectangular region with length $l_x=76$ and width $l_y=36$. The rectangular region is inside a ribbon with length $L_x=150$ and width $L_y=72$. Distribution of eigenvalues $\tilde{\epsilon}_k $ of the reduced density matrix of this region is plotted (there are also positive values of $\tilde{\epsilon}_k $, but the distribution is identical). The distribution follows the exponential curve $B[e^{(\tilde{\epsilon}-\tilde{\epsilon_0})^2/\delta^2}-1]$ (black solid line). The parameters are $\Gamma=0.3 t$, $n_{imp}=0.1$, and $U=0.5 t$ $(g=0.19)$. The number of disorder realization is $N_D \sim 50$.}
\label{calg0g3}
\end{center}
\end{figure}

When the operator $\tilde{h}$ is diagonalized, we get the following Hamiltonian matrix
\begin{eqnarray}
\tilde{h}_{ij}=\sum_{k}\psi^*_k(i)\psi_k(j)\tilde{\epsilon}_k,
\end{eqnarray}
where $\tilde{\epsilon}_k$ and $\psi_k(j)$ are eigenvalues and eigenstates of the ``Hamiltonian" $\tilde{h}$, respectively.
Note that this particular density matrix describes a Fermi gas at temperature $k_B T=1$.
The reduced density matrix of either spin-up or spin-down electrons is equal to
\begin{eqnarray}
\rho_{ij}= Tr(\rho c^{\dagger}_ic_j)=\sum_k \psi^*_k(i)\psi_k(j)\frac{1}{ e^{ \tilde{\epsilon}_k }+1 },
\end{eqnarray}
where $c_i=\sum_k \psi_k(i)a_k$, and $a_k$ is the electron destruction operator corresponding to the eigenstate $\psi_k(i)$. The distribution of the eigenvalues $\tilde{\epsilon}_k $ of $\tilde{h}$ is called the entanglement spectrum. (Note that $\tilde{\epsilon}_k $ are not the eigenenergies of the Hartree--Fock Hamiltonian.) The eigenvalues of a density matrix are given as
\begin{eqnarray}
\lambda_k=\frac{1}{ e^{ \tilde{\epsilon}_k }+1 }.
\end{eqnarray}
The values $\tilde{\epsilon}_k \approx 0$, corresponding to $\lambda_k\approx 1/2$, dominate the entanglement~\cite{KunY}.
The entanglement spectrum of the SPT phase of a disorder-free zigzag nanoribbon exhibits numerous nearly degenerate eigenvalues, thereby reflecting the presence of nearly degenerate edge states. Fig.~\ref{calg0g3} shows the HF entanglement spectrum of a disordered interacting graphene zigzag ribbon at half-filling. This entanglement spectrum is different from that of the disorder-free SPT phase of zigzag ribbons.
In the presence of disorder, the degenerate eigenvalues of the entanglement spectrum are split and exponentially distributed in a way similar to that of the DOS of edge states in TO zigzag ribbons, as shown in Fig. 1(b) in main article.  
In contrast, the DOS of disorder-free SPT phase of zigzag ribbons has a van Hove singularity, see  Fig. 1 in main article. 